\documentclass[conference]{IEEEtran}
\usepackage{cite}
\usepackage{url}
\usepackage{amsmath,amssymb,amsfonts}
\usepackage{amsmath,amssymb}
\usepackage{CJK}
\DeclareMathOperator{\argmin}{arg\,min} 

\usepackage{algorithm}
\usepackage{algorithmicx}
\usepackage{algpseudocode}
\usepackage{multirow}
\usepackage{graphicx}
\usepackage{textcomp}
\usepackage{xcolor}

\graphicspath{{fig/}}
\begin{document}

\title{Towards Privacy-Preserving Neural Architecture Search
}

\author{\IEEEauthorblockN{Fuyi Wang}
\IEEEauthorblockA{\textit{School of Information Technology} \\
\textit{Deakin University}\\
VIC 3216, Australia \\
wangfuyi@deakin.edu.au}
\and
\IEEEauthorblockN{Leo Yu Zhang}
\IEEEauthorblockA{\textit{School of Information Technology} \\
\textit{Deakin University}\\
VIC 3216, Australia \\
leo.zhang@deakin.edu.au}
\and
\IEEEauthorblockN{Lei Pan}
\IEEEauthorblockA{\textit{School of Information Technology} \\
\textit{Deakin University}\\
VIC 3216, Australia \\
l.pan@deakin.edu.au}
\and
\IEEEauthorblockN{Shengshan Hu}
\IEEEauthorblockA{\textit{School of Cyber Science and Engineering} \\
\textit{Huazhong University of Science and Technology}\\
Wuhan, China \\
hushengshan@hust.edu.cn}
\and
\IEEEauthorblockN{Robin Doss}
\IEEEauthorblockA{\textit{School of Information Technology} \\
\textit{Deakin University}\\
VIC 3216, Australia \\
robin.doss@deakin.edu.au}
}

\maketitle

\begin{abstract}
Machine learning promotes the continuous development of signal processing in various fields, including network traffic monitoring, EEG classification, face identification, and many more. However, massive user data collected for training deep learning models raises privacy concerns and increases the difficulty of manually adjusting the network structure. To address these issues, we propose a privacy-preserving neural architecture search (PP-NAS) framework based on secure multi-party computation to protect users' data and the model's parameters/hyper-parameters. PP-NAS outsources the NAS task to two non-colluding cloud servers for making full advantage of mixed protocols design.
Complement to the existing PP machine learning frameworks, we redesign the secure ReLU and Max-pooling garbled circuits for significantly better efficiency ($3 \sim 436$ times speed-up). 
We develop a new alternative to approximate the Softmax function over secret shares, which bypasses the limitation of approximating exponential operations in Softmax while improving accuracy. 
Extensive analyses and experiments demonstrate PP-NAS's superiority in security, efficiency, and accuracy.
\end{abstract}


\section{Introduction}
\label{sec:intro}


Signal processing based on neural networks is thriving due to its superior performance in various tasks, like face identification, network traffic classification, and incident prediction \cite{sun2019data,KemelmacherShlizerman2016TheMB}. The effectiveness of neural network algorithms depends on parameters and hyper-parameters to a large extent. In general, hyper-parameters are classified into two categories: training data-related parameters and network structure-related parameters. The automatic tuning of the training data-related parameters can employ traditional hyper-parameter optimization algorithms, including random search, grid search, Bayesian optimization, evolutionary algorithm, and many more. The automatic tuning of the network structure-related parameters is known as neural architecture search (NAS). NAS finds the optimal architecture to yield the best validation accuracy or other metrics like mean absolute deviation. Different from deep neural networks training, NAS solves an optimal parameter search problem in a high-dimensional space \cite{zoph2016neural,liu2018darts,wu2019fbnet}.

Without considering privacy, there are well-known works to achieve automatic search of neural architecture. Zoph et al.~\cite{zoph2016neural} proposed a combination of recurrent neural network (RNN) and policy-based reinforcement learning for rendering convolutional neural network (CNN) architectures. The RNN serves as the controller to output the hyper-parameters of a CNN, and a reinforcement learning component is used to supervise and improve the controller RNN. To update the controller RNN, the CNN must be trained from scratch, resulting in expensive computational costs. Liu et al.~\cite{liu2018darts} and Wu et al.~\cite{wu2019fbnet} defined an objective function that is differentiable against the hyper-parameters of the neural network architecture. This newly defined objective function successfully avoided reinforcement learning, resulting in saving computational expenses.


However, the enormous data used to train the model is likely to come from a variety of users and is sensitive in nature, violating data privacy laws like GDPR if left unprotected. 
To protect user privacy, privacy-preserving machine learning (PPML) based on secure multi-party computation (MPC) \cite{wagh2019securenn,mohassel2017secureml,mohassel2018aby3,khan2021blind} has been proposed and implemented. 
All these works protected user privacy by performing secure training over a fixed network architecture. However, merely avoiding leakage from users' training data does not equal user privacy preservation, let alone the fact that a fixed architecture does not guarantee the best performance in multiple tasks. 
The latest results in \cite{meng2019white,zanella2020analyzing} revealed that intermediate results like gradient updates leak users' private information.
Such leakage-abuse attacks are achieved by pivoting from the known neural network architecture during training. In this concern, it is imperative to protect structure-related parameters to ensure privacy. Therefore, the attacker who does not know about the neural network structure cannot reveal private information about training data even if the gradient update is leaked.


In response to the above-identified challenge, we investigate the privacy-preserving neural architecture search (PP-NAS) problem to improve trade-off between privacy and model performance.
The proposed PP-NAS utilizes the secret sharing technique to protect the private training data, weight parameters, and the architecture hyper-parameters against two non-colluding cloud servers. Moreover, we propose two sub-protocols to evaluate the ReLU activation and Max-pooling functions across shares with garbled circuits. 
We also present an alternative design of the private Softmax with a new approximating operation.
Mixing these elaborately designed sub-protocols with some existing secure protocols \cite{wagh2019securenn,mohassel2017secureml} enables the full-fledged solution of PP-NAS.
Extensive experiments demonstrate that PP-NAS performs similarly to the unprotected NAS, and the newly developed sub-protocols are more efficient than their PPML equivalents.

The rest of this paper is organized as follows. Sec.~\ref{sec:Preliminaries} introduces the preliminary knowledge used in the work. Sec.~\ref{sec:PP-NAS} presents the system model and our proposed hybrid protocol design for solving NAS. The experimental results are shown in Sec.~\ref{sec:ExperimentEvaluation} and the concluding remarks are drawn in Sec.~\ref{sec:Conclusion}.

\section{Preliminaries}
\label{sec:Preliminaries}

\subsection{NAS Search for CNN}
\label{ssec:NAS-CNN} 
The NAS paradigm aims to automatically search for the optimal network architecture that leads to the best validation accuracy or efficiency with time and resource constraints. 
It is consisted of three components: 1) defining the search space; 2) implementing the search strategy to sample the network; and 3) evaluating the performance of the searched network. More details are shown as follows.

\textbf{Search Space.} The search space clarifies the basic searchable units of the convolutional neural networks for the NAS algorithm.
{In a NAS implementation, each layer of a CNN is determined by searching from some basic units (cells or blocks): convolutional unit, activation unit, pooling unit, and linear/fully connected unit. 
With a proper evaluation, the unit with the best performance will be selected and concatenated in order to form the layer-by-layer architecture of the CNN.}

\textit{Convolutional Unit.} The convolutional unit is responsible for extracting the local features in an image through filtering with different convolution kernels, each of which is represented as a weight matrix. 
Assume there is a $n \times n$ convolution kernel $w$, with its bias being $b$, and a (partial) input image $x$ with size $n \times n$, the output convolution feature can be formulated as $y=\sum_{i=1}^{n} \sum_{j=1}^{n} w_{i, j} x_{i, j}+b$. 
Clearly, a convolutional unit can be formulated as the vector inner product between $w$ and $x$ (after adding a dummy element $x_{i,0} = 1$). 

\textit{Activation Unit.} The activation unit attached to each neuron is used to add nonlinear factors to enhance the feature expressiveness of the convolutional unit. 
Common activation functions include piece-wise linear and nonlinear functions with exponential characteristics, such as Sigmoid, Tanh, and ReLU. 
Numerous studies have shown that ReLU is a commonly used activation for CNN since it achieves the best performance in most applications, and it is recommended as the default activation function. 

\textit{Max-pooling Unit.} The pooling unit is used to aggressively reduce the dimensionality of each extracted feature map. 
This operation reduces the amount of computation and effectively avoids overfitting, because a Max-pooling unit outputs the maximum feature value within a sliding window.

\textit{Fully Connected Unit.} The fully connected unit is essentially the traditional multivariate linear regression. 
Similar to the convolutional unit, a fully connected unit can also be regarded as vector inner products. Given the weights $w$ and bias $b$ of the current layer, the output of $j$-th neuron is $y_j=\sum_{i} w_{i, j} x_{i}+b_{j}$. 

\textit{Remark.} Beside the searchable units for each layer, a classification CNN architecture always adopts its last layer as the Softmax to output a probability vector, which is called the confidence score of classification. And like literature works \cite{wagh2019securenn,agrawal2019quotient,mohassel2017secureml,mohassel2018aby3,demmler2015aby,xu2019cryptonn,khan2021blind}, we focus on NAS for classification CNN. 

\textbf{Search Strategy.} The search strategy defines how to find the optimal network structure, which is essentially an iterative optimization process for the architecture hyper-parameter. 
The existing mainstream strategy is a gradient-based optimization method, which can reduce the training times for effective search. Liu et al.~\cite{liu2018darts} proposed a  framework called Differentiable Architecture Search (DARTS), transforming NAS into an optimization problem in a continuous space and using the gradient descent method to search the neural architecture while obtaining the network weights simutaneously. Subsequently, improvement work emerged on the basis of DARTS, such as Progessive DARTS \cite{chen2019progressive}, Fair DARTS \cite{chu2020fair}. We also adopt this search strategy to design our scheme. 

\textbf{Performance Evaluation.} The goal of the search strategy is to find a neural network structure that maximizes some performance metrics, such as accuracy and efficiency. To guide the search process, the NAS algorithm needs to estimate the performance of a given neural network structure, which is called a performance evaluation strategy.

\subsection{Cryptographic Primitives}
\label{ssec:Cryptographic Primitives} 

To achieve PP-NAS, our scheme makes use of cryptographic tools, which are described in this section. 

\textbf{Additive Secret Sharing.} Suppose that there are two parties (i.e., cloud servers $\textnormal{S}0$ and $\textnormal{S}1$), an $l$-bit value $x$ is split into two random shares for each cloud server by additive secret sharing in the ring $\mathbb{Z}_{2^{l}}$, such that the sum of the two shares is equal to $x$. The two shares are denoted as $\left \langle x \right \rangle_{0}$ and $\left \langle x \right \rangle_{1}$ that are only known by the corresponding cloud server with $\left \langle x \right \rangle_{0}, \left \langle x \right \rangle_{1} \in \mathbb{Z}_{2^{l}}$, we have $\left \langle x \right \rangle_{0}+\left \langle x \right \rangle_{1}=x$ (mod $2^{l}$). To reconstruct $x$, donated as $Rec\left ( \cdot ,\cdot  \right )$, $\textnormal{S}0$ (or $\textnormal{S}1$) sends its shares $\left \langle x \right \rangle_{0}$ (or $\left \langle x \right \rangle_{1}$) to the other, who calculates $x=\left \langle x \right \rangle_{0}+\left \langle x \right \rangle_{1}$.

The basic operations on additive secret sharing are addition and multiplication. For \textit{Addition} $\textsf{Add}(\cdot, \cdot )$ of two shared values $\left \langle x \right \rangle$ and $\left \langle y \right \rangle$, $\textnormal{S}i$ locally computes $\left \langle z \right \rangle_{i}=\left \langle x \right \rangle_{i}+\left \langle y \right \rangle_{i}$, where $i\in \left \{ 0,1 \right \}$, obtaining $\left \langle z \right \rangle=\left \langle x \right \rangle+\left \langle y \right \rangle$. For \textit{Multiplication} $\textsf{Mult}(\cdot, \cdot )$ of two shared values $\left \langle x \right \rangle$ and $\left \langle y \right \rangle$ (i.e., $\left \langle z \right \rangle =\left \langle x \right \rangle \cdot \left \langle y \right \rangle$),  a pre-computed Beaver multiplication triplet 
\cite{beaver1991efficient} is used. In particular, given the triplet of the form $\left \langle c \right \rangle=\left \langle a \right \rangle \cdot \left \langle b \right \rangle$: $\textnormal{S}i$ computes $\left \langle s \right \rangle_{i}=\left \langle x \right \rangle_{i} - \left \langle a \right \rangle_{i}$ and $\left \langle t \right \rangle_{i}=\left \langle y \right \rangle_{i} - \left \langle b \right \rangle_{i}$ with $i\in \left \{ 0,1 \right \}$, both parties perform $Rec\left (\left \langle s \right \rangle_{0},\left \langle s \right \rangle_{1}  \right )$ and $Rec\left (\left \langle t \right \rangle_{0},\left \langle t \right \rangle_{1}  \right )$ to get $s$ and $t$. Then $\textnormal{S}i$  sets $\left \langle z \right \rangle_{i}=i\cdot s\cdot t +s \cdot \left \langle x \right \rangle_{i} +t \cdot \left \langle y \right \rangle_{i} +  \left \langle c \right \rangle_{i}$. Note that the sharing mentioned in the remainder of this paper is additive secret sharing.

\begin{figure*}[h]
  \centering
  \centerline{\includegraphics[width=16.5cm]{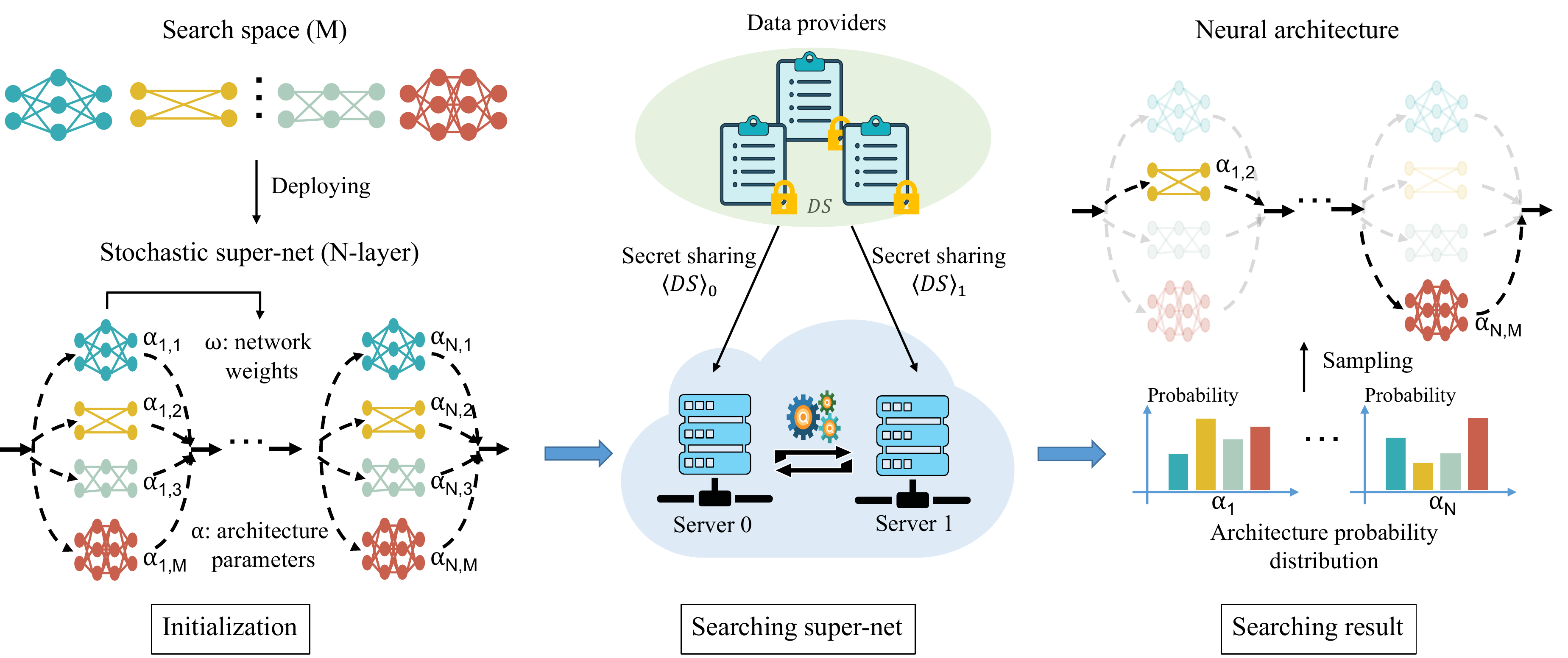}}
\caption{Overview of the privacy-preserving neural architecture search.}
\label{fig:PP-NAS} 
\end{figure*}

\textbf{Garbled Circuit.} Garbled circuit credited by Yao \cite{yao1982protocols} is a cryptographic protocol that enables two-party secure computation in which two mistrusting parties can jointly evaluate a function $f(x, y)$ over their private inputs $x$ and $y$ without the presence of a trusted third party, and nothing is learned about their inputs from the protocol other than the output. More detailed, one party, called the garbler, transforms the function $f(x, y)$ into a garbled Boolean circuit, which is a collection of garbled boolean gates. Then for each wire, the garbler specifies two random values as wire keys, corresponding to $0$ and $1$, and encrypts the output wire keys with all possible combinations of two input wire keys to create a garbled circuit table (GCT). The garbler sends the GCT to the other party, called the evaluator, along with its input key. The evaluator selects its input key by using oblivious transfer (OT), and decrypts only one element of the GCT with input keys from the garbler and himself. 

In this paper, we will use three simple garbled circuits, including addition (ADD), comparision (CMP), and multiplexer (MUX), to construct secure protocols. An ADD circuit is used to sum the two inputs. A CMP circuit realizes the function for comparison, that is, CMP$(x,y) =1$ if $x>y$; otherwise, it outputs $0$. 
The output of a MUX circuit depends on a selection bit, i.e., MUX$(x, y|b) = x$ if $b=0$; otherwise MUX$(x, y|b) = y$ for $b=1$. 

\section{Privacy-Preserving Neural Architecture Search}
\label{sec:PP-NAS}
\subsection{Overview of System Model}
\label{ssec:Overview}
In our PP-NAS system, there are two main parties: multiple data providers ($\textnormal{DPs}$) and two non-colluding cloud servers ($\textnormal{S}0$ and $\textnormal{S}1$). DPs hold datasets ($DS$) and will split them with the additive secret sharing into two shares, denoted by $\left \langle DS \right \rangle$ ($DS = \left \langle DS \right \rangle_0 + \left \langle DS \right \rangle_1$) for $\textnormal{S}0$ and $\textnormal{S}1$.
The two servers are considered as semi-honest, i.e., they faithfully execute the designated duties but attempt to learn additional information, such as $DS$ or the architecture, from their execution process. And they collaborate to search for network architecture hyper-parameters and train model weight parameters. 

As shown in Fig.~\ref{fig:PP-NAS}, our system consists of two phases: initialization and searching. In the initialization phase, we define a search space consisting of $M$ candidate units, the number of training epochs $E$, loss functions  $\mathcal{L_{A}} (\cdot)$ for model architecture and $\mathcal{L_{W}} (\cdot)$ model weights. Conceptually,  the ${M}$ units can establish an $N$-layer super-net. In the searching phase, the super-net is trained to obtain the architecture parameters $\mathcal{A}=\left \{ \alpha _{i,j} \right \} \left ( i=1, 2, \cdots ,N; j = 1, 2, \cdots, M \right )$. PP-NAS is  performed by the two cloud servers to jointly train the system architecture $\mathcal{A}$ and model weights $\mathcal{W}$ to minimize the loss function $\mathcal{L_{A}}$. In each epoch, the training is a bilevel optimization problem with $\mathcal{A}$ as the upper-level variable and $\mathcal{W}$ as the lower-level variable, defined as follows:
\begin{IEEEeqnarray}{rCL}
    \label{Eq:bilevelopt}
     \underset{\mathcal{A}}{\min} &\quad& \mathcal{L_{A}}\left (  \mathcal{W^*}\left( \mathcal{A} \right),\mathcal{A}\right )  \\
     \mathrm{s.t.} &\quad& \mathcal{W^*}\left ( \mathcal{A} \right )=\argmin_\mathcal{W}~\mathcal{L_{W}}\left ( \mathcal{W},\mathcal{A}   \right ). 
    \label{Eq:inneropt}
\end{IEEEeqnarray}
By approximating the optimal lower-level variable $\mathcal{W^*}$, Algorithm~\ref{alg:PP-NAS} presents the pseudocode for solving the above-nested optimization problem with gradient descent. 

\floatname{algorithm}{Algorithm}
\renewcommand{\algorithmicrequire}{\textbf{Input:}}
\renewcommand{\algorithmicensure}{\textbf{Output:}}
\begin{algorithm}[t]
    \caption{Privacy-preserving neural architecture search \label{alg:PP-NAS}}
    \begin{algorithmic}[1] 
        \Require $\textnormal{S}0: \left \langle DS \right \rangle_{0}$,  $\mathcal{L_{A}}\left ( \cdot  \right )$, $\mathcal{L_{W}}\left ( \cdot  \right )$, ${M}$, $N$
        
        $\textnormal{S}1: \left \langle DS \right \rangle_{1}$,  $\mathcal{L_{A}}\left ( \cdot  \right )$, $\mathcal{L_{W}}\left ( \cdot  \right )$, ${M}$, $N$
        
        \Ensure $\textnormal{S}0: \left \langle \mathcal{A} \right \rangle_{0}$, $\textnormal{S}1: \left \langle \mathcal{A} \right \rangle_{1}$
        \State Create a super-net with $N$ layers using $M$ units
        \State Initialize architecture $\mathcal{A}$ and $\mathcal{W}$
        \For{$i = 1 \to E$}
        \State $\left \langle \mathcal{A}_{i} \right \rangle := \left \langle \mathcal{A}_{i-1} \right \rangle - \eta _{\mathcal{A}}\left \langle \bigtriangledown _{\mathcal{A}}\mathcal{L_{A}}\left( \mathcal{W}_{i}, \mathcal{A}_{i-1} \right) \right\rangle$
        \State $\left \langle \mathcal{W}_{i} \right \rangle := \left \langle \mathcal{W}_{i-1} \right \rangle - \eta _{\mathcal{W}}\left \langle \bigtriangledown _{\mathcal{W}}\mathcal{L_{W}}\left( \mathcal{W}_{i-1},\mathcal{A}_{i-1}  \right) \right \rangle$
        \EndFor
        \State Obtain the final architecture $\left \langle \mathcal{A} \right \rangle$
    \end{algorithmic}
\end{algorithm}


\subsection{Our Design}
\label{ssec:OurDesign}


To enable $\textnormal{S}0$ and $\textnormal{S}1$ to perform Algorithm~\ref{alg:PP-NAS} with the shares, protocols for performing convolution/linear (or fully connected) unit, ReLU, Max-pooling, Softmax, and backpropagation are all mandatory. Since the convolution/liner (or fully connected) unit can be executed efficiently with the existing designs \cite{wagh2019securenn,mohassel2017secureml} that is  based on the Beaver multiplication triplet \cite{beaver1991efficient}, we focus on the other components in this paper. 

\begin{figure}[ht]
\begin{minipage}[b]{0.45\linewidth}
  \centering
  \centerline{\includegraphics[width=4.0cm]{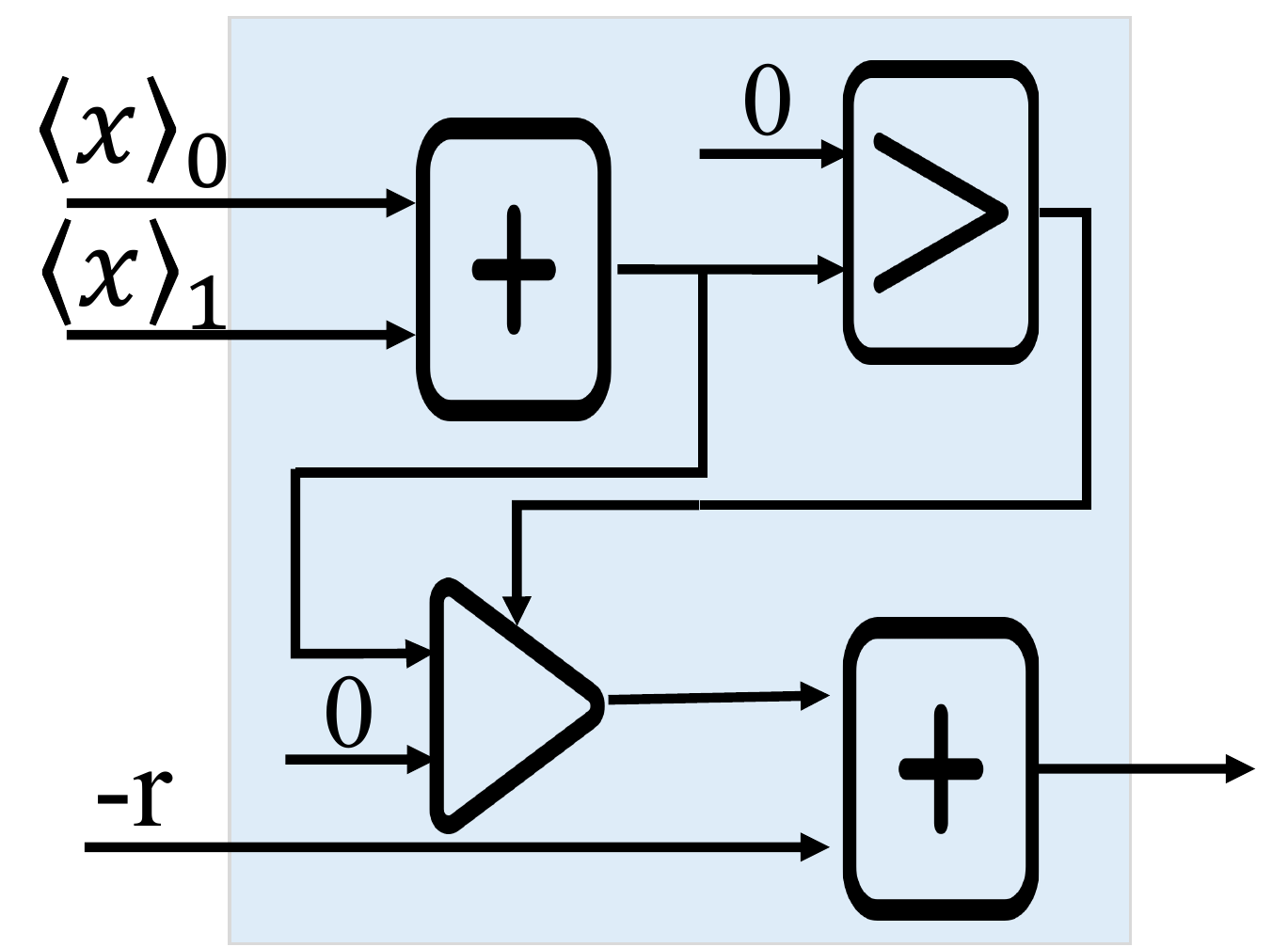}}
  \centerline{(a) ReLU circuit}\medskip
\end{minipage}
\hfill
\begin{minipage}[b]{0.45\linewidth}
  \centering
  \centerline{\includegraphics[width=3.2 cm]{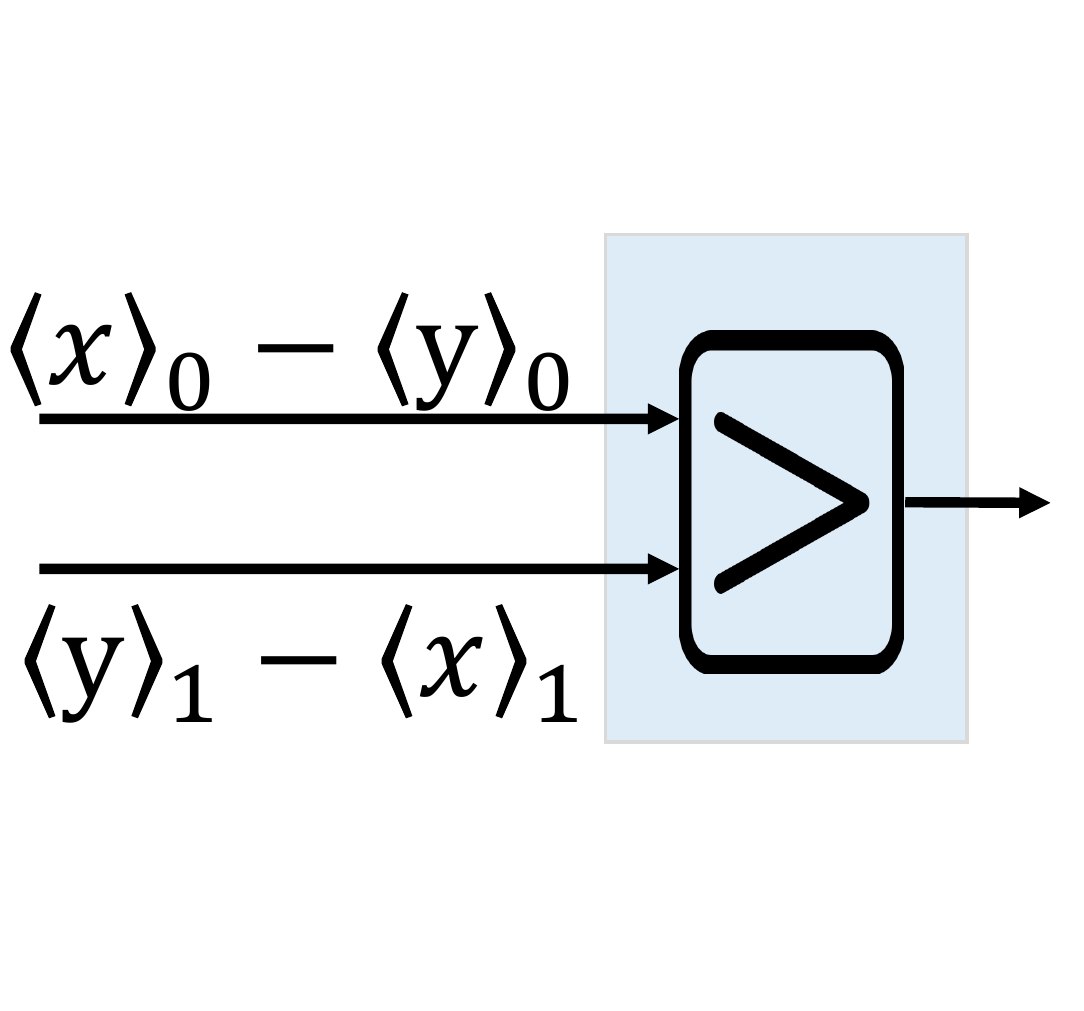}}
  \centerline{(b) Max-pooling circuit}\medskip
\end{minipage}
\caption{The structure of sub-protocols based on garbled circuits: ``$+$" denotes ADD, ``$>$" denotes CMP, and $\triangleright$ denotes MUX. Note that the output of MUX depends on CMP in (a).} 
\label{fig:circuit}
\end{figure}

\textbf{ReLU.} To achieve a ReLU activation, the existing work in \cite{wagh2019securenn} employs a series of privacy operations like most-significant bit computation and private comparison over shares. But we use garbled circuits to simplify the design.
Our solution involves four circuits, as shown in Fig.~\ref{fig:circuit}(a). The four circuits include one CMP, one MUX, and two ADDs. 
To securely evaluate ReLU$= \max(x, 0)$ over a shared $\left \langle x \right \rangle$ ($\textnormal{S}0$ has $\left \langle x \right \rangle_{0}$ and $\textnormal{S}1$ has $\left \langle x \right\rangle_{1}$), $\textnormal{S}0$ serves as the garbler generating the ReLU function circuit before sending it to the evaluator $\textnormal{S}1$. 
The garbler $\textnormal{S}0$ always holds $\left \langle x \right \rangle_{0}=r$, while the evaluator $\textnormal{S}1$ obtains $\left \langle x \right \rangle_{1}=x-r$ if $x > 0$ and $\left \langle x \right \rangle_{1}=-r$ otherwise. 

The ReLU garbled circuit is correct because, if $x>0$, 
we have CMP$(0, \textnormal{ADD}(\left\langle x \right\rangle_0, \left\langle x \right\rangle_{1})) = 0$ and ADD$( -r, \textnormal{MUX}(x, 0| 0) ) = x-r$. 
Similarly, the circuit is correct when $x<0$.

\textbf{Max-pooling.} Similar to ReLU, we use garbled circuits to implement the Max-pooling function. The building block for Max-pooling is to calculate $\max(x, y)$ over shares $\left \langle x \right \rangle_0$, $\left \langle y \right \rangle_0$, $\left \langle x \right \rangle_1$ and $\left \langle y \right \rangle_1$.
As shown in Fig.~\ref{fig:circuit}(b), our approach only uses one CMP circuit to achieve Max-pooling, which significantly simplifies the function design. Specifically, the circuit generator $\textnormal{S}0$ creates the CMP circuit with its input and sends it to the evaluator $\textnormal{S}1$. The evaluator $\textnormal{S}1$ then compares the received results with its input before sharing it with the garbler $\textnormal{S}0$. The numeric trick here for ensuring the correctness of $\max(x, y)$ over $\left \langle x \right \rangle$ and $\left \langle y  \right \rangle$ is because
\begin{IEEEeqnarray}{rCL}
&&(\left \langle x \right \rangle_0 -  \left \langle y \right \rangle_0) - (\left \langle y \right \rangle_1 -  \left \langle x \right \rangle_1) \nonumber\\ 
&=& (\left \langle x \right \rangle_0 +  \left \langle x \right \rangle_1) - (\left \langle y \right \rangle_0 +  \left \langle y \right \rangle_1). \nonumber
\end{IEEEeqnarray}

With the obtained binary comparison result, the two servers choose the shares corresponding to the larger one synchronously. If the evaluation result of the circuits is $1$, then $x > y$ holds, so both $\textnormal{S}0$ and $\textnormal{S}1$ select shares of $\left \langle x \right \rangle =\left \langle \max \right \rangle$; otherwise, $\left \langle y  \right \rangle = \left \langle \max \right \rangle $.

\textbf{Softmax.} It is challenging to design a private Softmax function due to its nonlinear nature \cite{khan2021blind}. As validated in Sec.~\ref{sec:PerformanceComparison}, the existing private Softmax will cause issues like slow convergence speed. In this regard, we present a privacy-friendly  approximation to replace the original implementation. 

Similar to the existing work in \cite{mohassel2017secureml}, the ReLU function is first applied to an input vector $\mathbf{ x }=[x_1, x_2 , \cdots, x_n]$ to ensure non-negative results.
To make the approximated Softmax smooth and scalable, we replace the original exponential function with the power function based on the following findings. 

With the fundamental fact that $\mathrm{exp}(x )$ can be approximated as
\begin{IEEEeqnarray}{rCL}
    \mathrm{exp}( x )&=&\lim_{n\rightarrow \infty }\left ( 1+{ x }/{n} \right )^{n} =\lim_{k\rightarrow \infty } \left ( 1+{ x }/{2^k} \right )^{2^k}, \nonumber
\end{IEEEeqnarray}
we define the notion $\widehat{\mathrm{exp}(x)}$ as 
{\small{
\begin{IEEEeqnarray}{rCL}
    \widehat{\mathrm{exp}( x )}  
    = \left(\frac{2^k}{e^{2^k}} \right)^{2^{k}} \mathrm{exp}( x )
    = \lim_{k\rightarrow \infty } \left( \frac{2^k+ x }{e^{2^k}} \right)^{2^{k}}
    \approx \left( \frac{ x }{e^{2^k}} \right)^{2^k}.
    \IEEEeqnarraynumspace \nonumber
\end{IEEEeqnarray}
}
}
\noindent We thus can compute the Softmax layer based on the above approximation, i.e., set $H=\left({2^k}/e^{2^k} \right)^{2^{k}}$, and use the fact
\begin{IEEEeqnarray}{rCL}
    H \cdot \widehat{\mathrm{exp}(x )}  
    = H^2 \mathrm{exp}(\left \langle x \right \rangle_{0} +\left\langle x \right\rangle_1 )
    =\widehat{\mathrm{exp}(\left \langle x \right \rangle_{0})} \widehat{\mathrm{exp}(\left \langle x \right \rangle_{1})} ,
    \IEEEeqnarraynumspace \nonumber
\end{IEEEeqnarray}
The Softmax of the shared vector $\left \langle\mathbf{x}\right \rangle =[\left \langle x_1 \right\rangle,  \cdots, \left\langle x_n \right\rangle]$ is 
{\small
\begin{IEEEeqnarray}{rCL}
    && \mathrm{Softmax}(\mathbf{x}) =  \left[  \widehat{\mathrm{exp}(x_1)}, \cdots, \widehat{\mathrm{exp}(x_n)} \right] \cdot ( \sum_{i=1}^{n} \widehat{\mathrm{exp}( x_i)} )^{-1}  
    \approx \IEEEeqnarraynumspace  \nonumber \\
    &  & \left[\cdots, \widehat{\mathrm{exp}(\left \langle x_i \right \rangle_{0})} \widehat{\mathrm{exp}(\left \langle x_i \right \rangle_{1})} , \cdots \right]  /
    ( \sum_{i=1}^{n} \widehat{\mathrm{exp}(\left \langle x_i \right \rangle_{0})} \widehat{\mathrm{exp}(\left \langle x_i \right \rangle_{1})} ).
    \IEEEeqnarraynumspace \nonumber
\end{IEEEeqnarray}
}
The above approximation is free in the sense that it directly operates over the shared $\left \langle\mathbf{x}\right \rangle$ and no expensive polynomial expansion of $\mathrm{exp}(\cdot)$ is needed \cite{khan2021blind}.
Moreover, it is clear the new approximation is smooth (and differentiable), resulting in the improvement of training efficiency \cite{ioffe2015batch}. 
As shown in Sec.~\ref{sec:PerformanceComparison}, by selecting an appropriate $k$, the proposed method is dramatically more accurate than the existing private Softmax.

\textbf{Private hyper-parameter/parameter optimization.} 
Even without considering privacy, solving Eq.~(\ref{Eq:bilevelopt}) with gradient descent is impractical because it involves the evaluation of $\bigtriangledown_{\mathcal{A}}\mathcal{L_{A}}\left( \mathcal{W^*}(\mathcal{A}), \mathcal{A}_{i-1}  \right)$, while $\mathcal{W^*}(\mathcal{A})$ is decided by the expensive inner optimization of Eq.~(\ref{Eq:inneropt}). As such, we follow the method\footnote{Note that this requires both loss functions $\mathcal{L_{A}}$ and $\mathcal{L_{W}}$ to be differentiable, and our Softmax approximation satisfies this requirement.} proposed by \cite{liu2018darts,wu2019fbnet} to approximate the needed gradient, i.e., 
\begin{IEEEeqnarray}{rCL}
    &&\bigtriangledown_{\mathcal{A}}\mathcal{L_{A}}\left( \mathcal{W^*}(\mathcal{A}), \mathcal{A}_{i-1}  \right) 
    \approx \bigtriangledown_{\mathcal{A}}\mathcal{L_{A}}\left( \mathcal{W}_i, \mathcal{A}_{i-1}  \right)
    \nonumber \\
    \approx&& \bigtriangledown_{\mathcal{A}} \mathcal{L_A}
    \left(\mathcal{W}_{i-1} -\eta_{\mathcal{W}}  \bigtriangledown_{\mathcal{W}}\mathcal{L_{W}}\left( \mathcal{A}_{i-1}, \mathcal{W}_{i-1}  \right), \mathcal{A}_{i-1} \right). \nonumber
\end{IEEEeqnarray}
%
With this prior, the two cloud servers $\textnormal{S}0$ and $\textnormal{S}1$ can update the {architecture parameters} with their shares according to
\begin{IEEEeqnarray}{rCL}
    \left \langle \mathcal{A}_{i} \right \rangle = \left \langle \mathcal{A}_{i-1} \right \rangle - {\eta_{\mathcal{A}}}/{\left | B \right |} \left \langle \bigtriangledown _{\mathcal{A}}\mathcal{L_{A}}
    \left( \mathcal{W}_{i}, \mathcal{A}_{i-1} \right ) \right \rangle, \nonumber
\end{IEEEeqnarray}
where $\left | B \right |$ donates the batch size. 
Similarly, $\mathcal{W}$ is optimized for the fixed $\mathcal{A}_{i}$ over the shares, i.e.,
\begin{IEEEeqnarray}{rCL}
    \left \langle \mathcal{W}_i \right \rangle = \left \langle \mathcal{W}_{i-1} \right \rangle - {\eta_{\mathcal{W}}}/{\left | B \right |}\left \langle \bigtriangledown _{\mathcal{W}}\mathcal{L_{W}}\left ( \mathcal{A}_{i-1}, \mathcal{W}_{i-1}  \right ) \right \rangle. \nonumber
\end{IEEEeqnarray}

\subsection{Security Analysis}
\label{sec:SecurityAnalysis}
Recall that the cloud servers are modeled as semi-honest in Sec.~\ref{ssec:Overview}, i.e., they strictly follow the implementation of our protocols but attempt to infer the secret information while executing the protocol.
When performing NAS searching, the dataset always appears to be random numbers under additive secret sharing, and is secured against cloud servers (and outsider attackers).
In addition, all the intermediate computation results about the architecture and the weight of the network are split between the two cloud servers (once again as random numbers), so no single server alone can extract the meaningful information from intermediate results. 

In the garbled circuit, the garbling scheme satisfies the standard security properties formalized in \cite{bellare2012foundations}.
Readers may refer to \cite{lindell2009proof} for a detailed description and proof of security against semi-honest adversaries. And our secure ReLU and Max-pooling sub-protocols are based on a garbling scheme over shared data. In a nutshell, our design achieves the security goal when performing neural architecture search. 


\section{Experiment Evaluation}
\label{sec:ExperimentEvaluation}
\subsection{Experimental Setup}
\label{sec:ExperimentalSetup} 
To verify the effectiveness and efficiency of PP-NAS, we evaluate our approach over the MNIST dataset 
\cite{mnistweb}, where 60,000 samples are used for training and 10,000 samples for testing. The batch size is set to $128$. The search space in our system is the same as the one in  \cite{liu2018darts}. The two parties in our system are implemented on two separate cloud servers with NVIDIA RTX 2080Ti GPU. Our experiments are conducted using a software framework CrypTen \cite{crypten2020} built with Python 3.7 and Pytorch 1.9. 
\subsection{Performance Comparison}
\label{sec:PerformanceComparison}

\textbf{Supporting Protocols.}
For model training, we propose two supporting protocols (i.e., ReLU and Max-pooling), which are constructed by FastGC \cite{huang2011faster} with the free-XOR technique \cite{kolesnikov2008improved} and point-and-permute optimizations. 
We test ReLU and the $2\times 2$ Max-pooling function with shares generated from a large enough ring (defined in \ref{ssec:Cryptographic Primitives}) that retrains the plaintext precision (i.e., double precision floating-point arithmetic) \cite{mohassel2017secureml,wagh2019securenn}, and the results are tabulated in Table ~\ref{tab:SubprotocolComparison} by comparing to two state-of-the-art works: CrypTen \cite{crypten2020} and SecureNN \cite{wagh2019securenn}.

Observing this table, it clear that the simple Max-pooling circuit brings an incredible advantage of communication cost ($1.04$Kb) and running time overhead ($0.000183$s), respectively. Clearly, our method outperforms all existing methods in running time (roughly $68\times$ faster than SecureNN and $436 \times$ faster than CrypTen). Moreover, our communication cost is close to that of SecureNN, which improves on CrypTen by about 7$\times$.

Our ReLU function has an advantage in running time, roughly 3$\times$ faster than SecureNN and 19$\times$ faster than CrypTen. Due to use a slightly complicated circuit to achieve the ReLU function, our method incurs a moderate communication overhead. This trade-off is acceptable. This is because when the communication costs are within a certain range, the computation costs are normally a more important measure of performance, which should be 
reduced as much as possible.


\begin{figure}[t]
  \centering
  \centerline{\includegraphics[width=9cm]{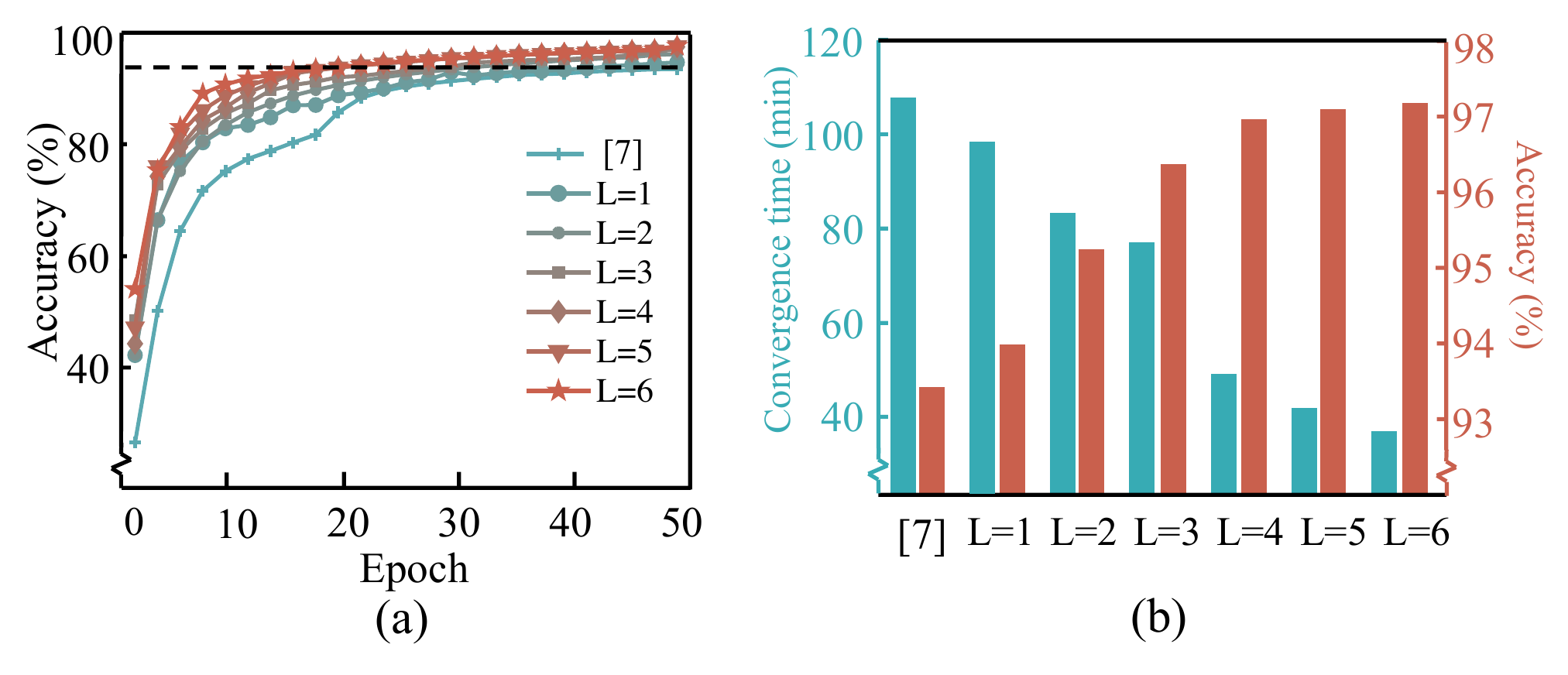}}
\caption{Performance comparison of private Softmax alternatives. Note that the convergence time refers to the training time when the accuracy first reaches $93.4$\%.} 
\label{fig:SoftmaxResult} 
\end{figure}

\begin{table}[!t]
\centering
\caption{The comparison of the communication and running time among our supporting protocols and previous works.}
\label{tab:SubprotocolComparison}
\resizebox{1\linewidth}{!}{
\begin{tabular}{c|c|c|c|c}
\hline
\multirow{2}{*}{\textbf{Models}} & 
\multicolumn{2}{c|}{\textbf{Comm.~(Kb)}} & 
\multicolumn{2}{c}{\textbf{Time (s)}} \\ \cline{2-5} 
 & \multicolumn{1}{l|}{ReLU} & \multicolumn{1}{l|}{Max-pooling} & \multicolumn{1}{l|}{ReLU} & \multicolumn{1}{l}{Max-pooling} \\ \hline
CrypTen \cite{crypten2020}   & 2.18  & 7.03  & 0.018653 & 0.079837     \\ 
SecureNN  \cite{wagh2019securenn} & 0.46        & 1.47    & 0.003077   & 0.012534     \\
Ours   &  6.16        & 1.04       & 0.000948        & 0.000183   \\ \hline
\end{tabular}
}
\end{table}

\textbf{Approximation of Softmax.} To visually evaluate our approximated Softmax function, we consider the \textit{Network A} in SecureNN. It is composed of two fully connected layers with ReLU activation and a Softmax layer. The model is trained by PyTorch with the alternative Softmax function \cite{mohassel2017secureml} $f_1\left ( x_{i} \right )=\frac{\mathrm{ReLU}\left ( x_{i}  \right )}{\sum \mathrm{ReLU}\left ( x_{i}  \right )}$ and our approximated Softmax $f_2$ in Sec.~\ref{ssec:OurDesign} with $L=2^k$ set to $1, 2, 3,4,5,6$, respectively. 

Although these two alternatives both guarantee that the network's final output is a probability distribution without leaking privacy, as shown in Fig.~\ref{fig:SoftmaxResult}, the model converges faster with better accuracy when using $f_2$. 
As shown in Fig.~\ref{fig:SoftmaxResult}(a), with different choice of $L$, our method consistently outperforms $f_1$ in \cite{mohassel2017secureml} regarding training accuracy, and the best accuracy $f_1$ can achieve is $93.4$\%.  Fig.~\ref{fig:SoftmaxResult}(b) plots the training time of $f_2$ when the accuracy first reaches $93.4$\%, and the saving in time is very clear for all choices of $L$. In addition, Fig.~\ref{fig:SoftmaxResult}(b) shows that, with $f_2$, the \textit{Network A} shows an upward trend with the increase of $L$, and can reach $97.2$\% accuracy with $L=6$. 
All these findings support our argument in Sec.~\ref{ssec:OurDesign}: smooth and differentiable private Softmax approximation improves both training accuracy and efficiency.
Theoretically, before reaching the performance of the plain \textit{Network A}, larger value of $L$ (e.g., $L>6$) should lead to even better result (i.e., less training time and higher accuracy). However, due to the exponentiation nature of $f_2$, larger $L$ leads to representation overflow. 

\begin{figure}[t]
  \centering
  \centerline{\includegraphics[width=9cm]{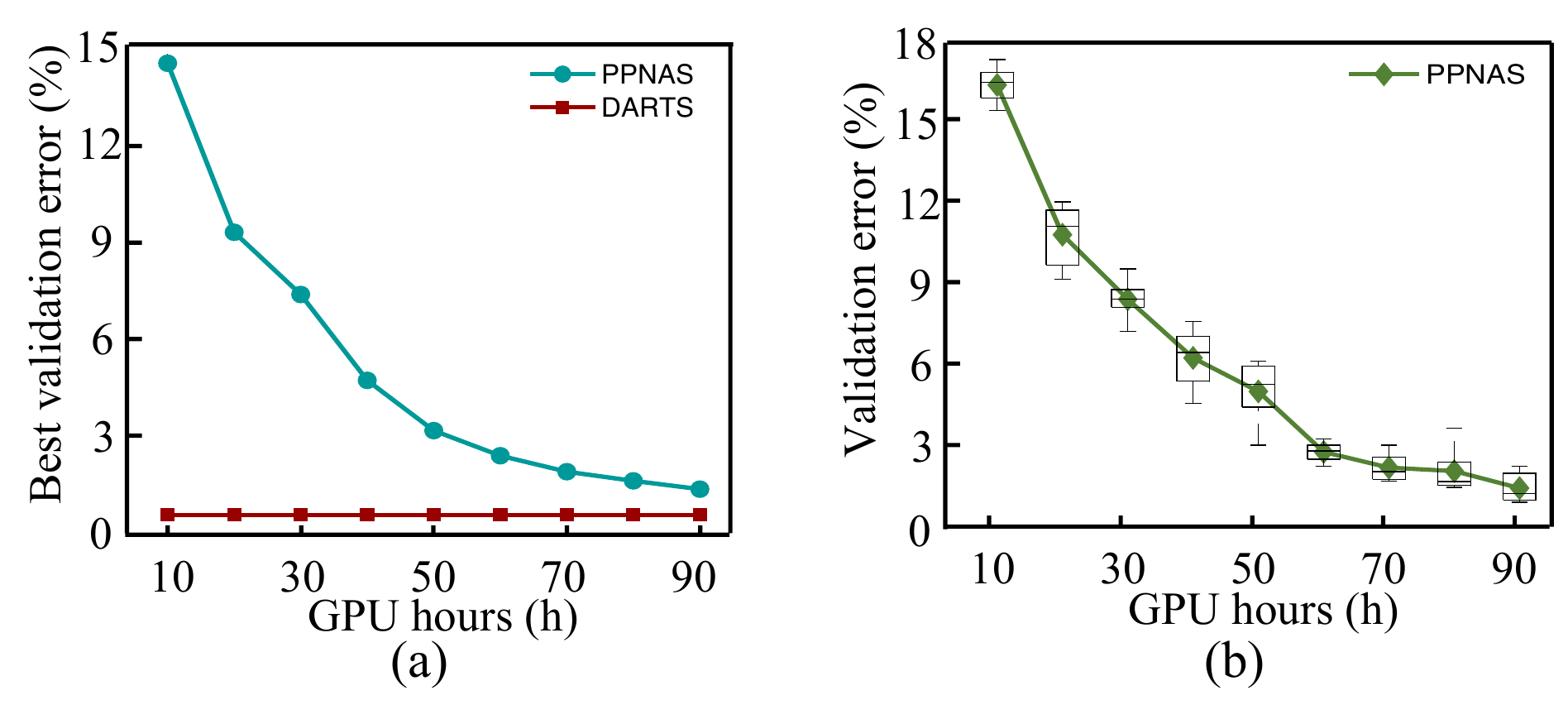}}
\caption{Best validation error and average validation error for MNIST.} 
\label{fig:testResult} 
\end{figure}
\textbf{Overall Performance.} We regard DARTS in \cite{liu2018darts} without privacy consideration as our baseline, and plot the model validation error with GPU hour in Fig.~\ref{fig:testResult}. It is observed from this figure that the best validation error has a steady downward trend for PP-NAS and the model accuracy in PP-NAS can reach to $98.59$\%. As usual, the use of privacy-protection technique brings an increase in training time and a slight loss of accuracy. 
We further evaluate PP-NAS with several state-of-the-art privacy-preserving machine learning schemes \cite{xu2019cryptonn,wagh2019securenn,abadi2016deep}. Here, $L=2^k=4$ 
is used. 
The accuracy from the different secure models is listed in Table \ref{tab:TotalSystem}, which again indicates that PP-NAS maintain good functionality with ensured privacy. Note that there is a security/functionality trade-off for the work \cite{abadi2016deep} using differential privacy, and we only report the best result (i.e., worst privacy) here. 


\begin{table}[t]
\centering
\caption{Comparison of the model accuracy among our design and other state-of-the-art secure training methods.}
\label{tab:TotalSystem}
\resizebox{1\linewidth}{!}{
\begin{tabular}{c|c|c|c|c}
\hline
\textbf{Dataset} & \textbf{PP-NAS} & \textbf{CryptoNN \cite{xu2019cryptonn}} & \textbf{SecureNN \cite{wagh2019securenn}} & \textbf{Abadi et al.~\cite{abadi2016deep}} \\ \hline
MNIST   &  98.59\%    & 95.94\%  & 93.4\%   & 97.52\%  \\ \hline
\end{tabular}
}
\end{table}

\section{Conclusion}
\label{sec:Conclusion}
\noindent In this paper, we present a neural architecture search scheme based on secure MPC. All the hyper-parameters/parameters are protected against the semi-honest cloud servers. Using the garbled circuit, we designed two sub-protocols to significantly reduce the run time for practical deployment. An approximated Softmax has been proposed to replace the exponential operation, which can approximate the Softmax function with ensured privacy. Compared with the literature, our proposed protocol finds a better neural architecture with higher accuracy. We consider the application of the proposed method to some privacy-critical applications as future work. 




\bibliographystyle{IEEEbib}
\bibliography{main}

\end{document}